\documentclass[12pt]{article}
\usepackage{epsfig}
\usepackage{latexsym}
\title{Classical Physics and Quantum Loops}
\author{Barry R. Holstein$^{a,b}$ and John F. Donoghue$^a$\\
$^a$ Department of Physics-LGRT\\
University of Massachusetts\\
Amherst, MA  01003\\
$^b$ Theory Group\\
Thomas Jefferson National Accelerator Laboratory\\
12000 Jefferson Ave.\\
Newport News, VA 23606}
\begin{document}
\begin{titlepage}
\maketitle
\begin{abstract}
The standard picture of the loop expansion associates a factor of $\hbar$ with each
loop, suggesting that the tree diagrams are to be associated with classical physics,
while loop effects are quantum mechanical in nature.  We discuss examples wherein
classical effects arise from loop contributions and display the relationship between
the classical terms and the long range effects of massless particles.
\end{abstract}
\end{titlepage}
\section{Introduction}

It is commonly stated that the loop expansion in quantum field theory is equivalent
to an expansion in $\hbar$. Although this is mentioned in several field theory
textbooks, we have not found a fully compelling proof of this statement. Indeed, no
compelling proof is possible because the statement is not true in general. In this
paper we describe several exceptions - cases where classical effects are found
within one loop diagrams - and discuss what goes wrong with purported ``proofs''.

 Most
physicists performing quantum mechanical calculations eschew
keeping track of factors of $\hbar$, and use units wherein $\hbar$
is set to unity---only when numerical results are needed are these
factors restored.  However, use of this procedure can cloak the
difference between classical and quantum mechanical effects, since
the former are distinguished from the latter merely by the absence
of factors of $\hbar$.  This is also the practice in many field
theory texts, but there is often a discussion in such works about
a one to one connection between the number of loops and the
factors of $\hbar$\cite{ft}.  The argument used in order to make
this connection is a simple one, and is worth outlining here:  In
calculating a typical Feynman diagram, the presence of a vertex
arises from the expansion of

$$\exp {i\over \hbar}\int{\cal L}_{int}(\phi_{in})d^4x$$
and so carries with it a factor of $\hbar^{-1}$.  On the
other
hand the field commutation relations

$$[\phi(\vec{x}),\pi(\vec{y}]=i\hbar\delta^3(\vec{x}-
\vec{y})$$
lead to a factor of $\hbar$ in each propagator

$$<0|T(\phi(x)\phi(y))|0>=\int{d^4k\over
(2\pi)^4}{i\hbar e^{ik(x-y)}\over k^2-\frac{m^2}{\hbar^2}-i\epsilon}$$ The counting
of factors of $\hbar$ then involves calculating the number of vertices and
propagators in a given diagram. For a diagram with V vertices and I internal lines
the number of independent momenta is $L=I-V+1$ and corresponds to the number of
loops. Associating a factor of $\hbar^{-1}$ for the V vertices and $\hbar^{+1}$ for
the I propagators yields an overall factor
$$\hbar^{I-V+1}=\hbar^{L}$$
which is the origin of the claim that the loop expansion coincides
with an expansion in $\hbar$.  We shall demonstrate in the next
section, however, that this assertion in not valid.

\section{A counterexample}

Let us give an example where one obtains classical results from a
one-loop calculation. This example is chosen because it is easy to
identify the classical and quantum effects. We describe the
one-loop QED calculation of the matrix element of the energy-
momentum tensor between initial and final plane wave
states\cite{db1}.  For simplicity, we shall discuss below the case
wherein these states are spinless, but the calculation was
performed also for spin 1/2 systems and the results are the same.

The basic structure of the matrix element is given by
\begin{equation}
<p_2|T_{\mu\nu}(0)|p_1>={1\over \sqrt{4E_2E_1}}[2P_\mu P_\nu F_1(q^2)+(q_\mu
q_\nu-\eta_{\mu\nu} q^2)F_2(q^2)]
\end{equation}
where  $F_1(q^2),\, F_2(q^2)$ are form factors, to be determined. In lowest order
the energy-momentum tensor form factors are
\begin{equation}
F_1(q^2)=1,\,\,F_2(q^2)=-{1\over 2},\label{eq:ff}
\end{equation}
but these simple results are modified by loops, and the form
factors will receive corrections of order $e^2$ at one loop order.
One can evaluate these modifications using the diagrams shown in
Figure 1, and the results are found to be\cite{db1}
\begin{eqnarray}
F_1(q^2)=1+{e^2\over 16\pi^2}{q^2\over m^2}\left({3\over 4}{m\pi^2\over
\sqrt{-q^2}}-{8\over 3}+2\log{-q^2\over
m^2}-{4\over 3}\log{\lambda\over m}\right)+\ldots\nonumber\\
F_2(q^2)=-{1\over 2}+{e^2\over 16\pi^2} \left({m\pi^2\over 2\sqrt{-q^2}}-\Omega-
{26\over 9}+{4\over 3}\log{-q^2\over m^2}\right)+\ldots
\end{eqnarray}
where
\begin{equation}
\Omega={2\over \epsilon}-\gamma-\log{m^2\over 4\pi\mu^2}
\end{equation}
The factors of $\hbar$ will be inserted in the discussion of the next section.

\begin{figure}
\begin{center}
\epsfig{file=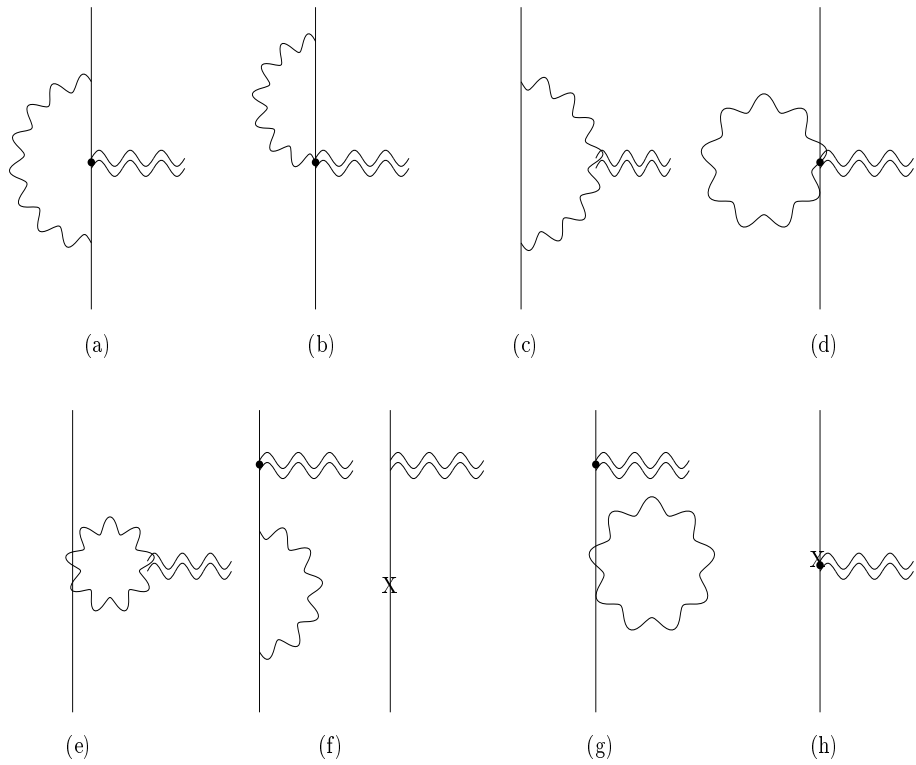,height=10cm,width=12cm} \caption{Feynman diagrams for
spin 0 radiative corrections to $T_{\mu\nu}$.}
\end{center}
\end{figure}

It is easiest to separate the classical and quantum effects by going to coordinate
space via a Fourier transform. The key terms are those that have a nonanalytic
structure such as $\sqrt{-q^2/m^2}$ and $q^2\ln-q^2$. These both arise only from
those diagrams where the energy momentum tensor couples to the photon lines. In
particular, the square root term comes uniquely from Figure 1c. We will see that the
square root turns into a well known classical correction while the logarithm
generates a quantum correction. Specifically we take the transform
\begin{equation}
T_{\mu\nu}(\vec{r}) = \int {d^3q\over (2\pi)^3}e^{i\vec{q}\cdot\vec{r}}
T_{\mu\nu}(\vec{q})
\end{equation}
Using
$$\int{d^3q\over (2\pi)^3}e^{i\vec{q}\cdot\vec{r}}|\vec{q}|=-{1\over
\pi^2r^4}$$
as well as
$$\int{d^3q\over (2\pi)^3}e^{i\vec{q}\cdot\vec{r}}\vec{q}^2\log \vec{q}^2
={3\over \pi r^5}$$
and including powers of $\hbar$ in the result,
we find
\begin{eqnarray}
T_{00}(\vec{r})&=&m\delta^3(\vec{r})+{e^2\over 32\pi^2 r^4}
-{e^2\hbar\over 4\pi^3mr^5}+\ldots\nonumber\\
T_{0i}(\vec{r})&=&0\nonumber\\
T_{ij}(\vec{r})&=& -{e^2\over 16\pi^2 r^4}\left({r_ir_j\over
r^2}-{1\over 2}\delta_{ij}\right)-{e^2\hbar\over
16\pi^3mr^5}(3\delta_{ij}-5{r_ir_j\over r^2})+\ldots\label{eq:tm}
\end{eqnarray}
We see then that Eq. \ref{eq:tm} includes both corrections which
are independent of $\hbar$ as well as pieces which are linear in
this quantity.

The interpretation of the classical terms is clear.  Since the
energy-momentum tensor for the electromagnetic field has the
form\cite{ja}
\begin{equation}
T_{\mu\nu}^{EM}=-F_{\mu\lambda}F_\nu{}^\lambda+{1\over
4}\eta_{\mu\nu}F_{\lambda\delta} F^{\lambda\delta}
\end{equation}
and, for a simple point charge, we have
$$\mathbf{E}=\frac{e}{4\pi r^2}\mathbf{\hat{r}}$$
we determine
\begin{eqnarray}
T_{00}^{EM}(\vec{r})&=&{1\over 2}E^2={e^2\over 32\pi^2
r^4}\nonumber\\
T_{0i}^{EM}(\vec{r})&=&0\\
T_{ij}^{EM}(\vec{r})&=&-E_iE_j+{1\over 2}\delta_{ij}E^2=-{e^2\over 16\pi^2
r^4}\left({r_ir_j\over r^2}-{1\over 2}\delta_{ij}\right)
\end{eqnarray}
which agree exactly with the component of Eq. \ref{eq:tm} which
falls as $1/r^4$.  Despite arising from a loop calculation then
this is a {\it classical} effect, due to the feature that the
energy momentum tensor can couple to the electric field
surrounding the particle as well as to the particle directly. At
tree level, the energy momentum tensor represents only that of the
charged particle itself.  However, the charged particle has an
associated classical electric field and that field also carries
energy-momentum. The one loop diagrams where the energy momentum
tensor couples to the photon lines correspond to the process
whereby the charged particle generates the electric field, which
is in turn and measured by the energy momentum tensor. From this
point of view, it is not surprising that the calculation yields a
classical term - there is energy in the classical field at this
order in $e^2$ and a calculation at order $e^2$ must be capable of
uncovering it.

Of course, the full loop calculation also contains additional physics, the leading
piece of which is quantum mechanical in nature and falls as
$\hbar/mr^5$.\footnote{The form of these terms can be understood in a handwaving
fashion from the feature that while the distance $r$ between a source and test
particle is well defined classically, at the quantum level there are fluctuations of
order the Compton wavelength
$$r\longrightarrow r+\delta r$$
with $\delta r\sim \hbar/m$. When expanded via
$$1/ (r+\delta r)^4\sim{1\over r^4}-4{\delta r\over
r^5}={1\over r^4}-{4\hbar\over mr^5}$$ we see that the form of
such corrections is as found in the loop calculation.  That such
Compton wavelength corrections are quantum mechanical in nature,
as can be seen from the explicit factor of $\hbar$.} So we see
that the one loop diagram contains both classical and quantum
physics.

\section{What Went Wrong?}

The argument that the loop expansion is equivalent to an expansion
in $\hbar$ clearly failed in the above calculation, and in this
section we shall examine this failure in more detail.

One loophole to the original argument is visible in the
propagator, which contains $\hbar$ in more than one location. When
the propagator written in terms of an integral over the
wavenumber, the mass carries an inverse factor of $\hbar$. This is
because the Klein-Gordon equation reads
$$ (\Box + \frac{m^2}{\hbar^2}) \phi(x) =0 $$
when $\hbar$ is made visible.  This means that the counting of
$\hbar$ from the vertices and the propagator is incomplete---one
also needs to know how the mass enters the result, because there
are factors of $\hbar$ attached therein also.

In the previously discussed loop calculation of the formfactors of
the energy momentum tensor, we can display the factors of $\hbar$
in momentum space. Returning $\hbar$ to the formula for $F_1$ we
find (we continue to use $c=1$)
\begin{eqnarray}F_1(k)&=& 1+{e^2\over 16\pi^2\hbar}{\hbar^2 k^2\over
m^2}\left({3\over 4}{m\pi^2\over \sqrt{-\hbar^2k^2}}-{8\over
3}+2\log{-\hbar^2k^2\over m^2}-{4\over 3}\log{\lambda\over
m}\right)+\ldots  \nonumber \\
&=& 1 + \frac{3e^2\sqrt{-k^2}}{64m}+\frac{\hbar e^2 k^2}{16\pi^2 m^2}\left(-{8\over
3}+2\log{-\hbar^2k^2\over m^2}-{4\over 3}\log{\lambda\over m}\right)+\ldots
\end{eqnarray}
Here we have written the momentum in terms of the wavenumber
$q=\hbar k$, and we note that $e^2/\hbar$ is dimensionless in
Gaussian units(with $c=1$). It is easy to see then that the
coefficient of the square root nonanalytic behavior is independent
of $\hbar$, while the logarithmic term has one power of $\hbar$
remaining. This is fully consistent with the coordinate space
analysis of the previous section and illustrates the feature that
terms which carry different powers of the momentum and mass can
have different factors of $\hbar$.

We see then that the one loop result carries different powers of $\hbar$ because it
contains different powers of the factor $q^2/m^2$. Moreover, we can be more precise.
With the general expectation of one factor of $\hbar$ at one loop, there is a
specific combination of the mass and momentum that eliminates $\hbar$ in order to
produce a classical result.  In order to remove one power of $\hbar$ requires a
factor of
\begin{equation}
\sqrt{\frac{m^2}{-q^2}}= \frac{m}{\hbar\sqrt{-k^2}}
\end{equation}
This is a nonanalytic term which is generated only by the propagation
of massless particles. The emergence of the power of $\hbar^{-1}$ involves an
interplay between the massive particle (whose mass carries the factor of $\hbar$)
and the massless one (which generates the required nonanalytic form). This result
suggests that one can generate classical results from one loop processes in the
presence of massless particles, which have long range propagation and therefore
generate the required nonanalytic momentum behavior.

\section{Additional Examples}

In this section we describe other situations where classical
results are found in one loop calculations.  All involve couplings
to massless particles.

The calculation of the energy momentum tensor can be extended to
include graviton loops as has been done in Ref.\cite{db2}. Here
there exists a superficial difference in that the gravitational
coupling constant carries a mass dimension and the one loop result
involves the Newtonian gravitational constant $G_N$.  This feature
might be thought to change the counting in $\hbar$, but it does
not.  Again the important diagrams are those in which the energy
momentum vertex couples to the graviton line. The resulting
(spinless) form factors were found to be

\begin{eqnarray}
F_1(q^2)&=&1+{Gq^2\over \pi}(
{1\over 16}{\pi^2 m\over \sqrt{-q^2}}-{3\over 4}\log{-q^2\over m^2})+\ldots\nonumber\\
F_2(q^2)&=&-{1\over 2}+{Gm^2\over \pi}({7\over 8}{\pi^2m\over
\sqrt{-q^2}}-2\log{-q^2\over m^2})+\ldots
\end{eqnarray}
corresponding to a co-ordinate space energy-momentum tensor:

\begin{eqnarray}
T_{00}(\vec{r})&=&m \delta^3(r)-{3Gm^2\over 8\pi
r^4}-{3Gm\hbar \over 4\pi^2 r^5}+\ldots\nonumber\\
T_{0i}(\vec{r})&=&0\nonumber\\
T_{ij}(\vec{r})&=&-{7Gm^2\over 4\pi r^4}\left({r_ir_j\over
r^2}-{1\over 2}\delta_{ij}\right)+{Gm\hbar\over
2\pi^2r^5}\left(9\delta_{ij}-15{r_ir_j\over
r^2}\right)+\ldots\label{eq:h1}
\end{eqnarray}

This result can be compared with that arising from the classical
energy-momentum pseudo-tensor for the gravitational field\cite{we}

\begin{eqnarray}
8\pi GT_{\mu\nu}^{\rm grav}&=&-{1\over
2}h^{(1)\lambda\kappa}[
\partial_\mu\partial_\nu
h^{(1)}_{\lambda\kappa}+\partial_\lambda\partial_\kappa
h^{(1)}_{\mu\nu}\nonumber\\
&-&\partial_\kappa\left(\partial_\nu
h^{(1)}_{\mu\lambda}+\partial_\mu
h^{(1)}_{\nu\lambda}\right)]\nonumber\\
&-&{1\over 2}\partial_\lambda
h^{(1)}_{\sigma\nu}\partial^\lambda
h^{(1)\sigma}{}_\mu +{1\over 2}\partial_\lambda
h^{(1)}_{\sigma\nu}\partial^\sigma h^{(1)\lambda}{}_\mu -
{1\over
4}\partial_\nu h^{(1)}_{\sigma\lambda}\partial_\mu
h^{(1)\sigma\lambda}\nonumber\\
&-&{1\over 4}\eta_{\mu\nu}(\partial_\lambda
h^{(1)}_{\sigma\chi}
\partial^\sigma h^{(1)\lambda\chi}
-{3\over 2}\partial_\lambda
h^{(1)}_{\sigma\chi}\partial^\lambda
h^{(1)\sigma\chi}) -{1\over 4}h^{(1)}_{\mu\nu}\Box
h^{(1)}\nonumber\\
&+&{1\over 2}\eta_{\mu\nu}h^{(1)\alpha\beta}\Box h^{(1)}
_{\alpha\beta}\label{eq:h2}
\end{eqnarray}
Using the lowest order solution
\begin{equation}
h^{(1)}_{\mu\nu}(r)=-\delta_{\mu\nu}{2Gm\over r}
\end{equation}
the $1/r^4$ components of Eqs. \ref{eq:h1} and \ref{eq:h2} are
seen to agree.  Equivalently, the expression of the energy
momentum tensor can be used to calculate the metric around the
particle\cite{db2}. Doing so yields the nonlinear classical
corrections to order $G^2$ in the Schwarzschild metric (in
harmonic gauge)
\begin{eqnarray}
g_{00}&=&1-2{Gm\over r}
+2{G^2m^2\over r^2}+\ldots\nonumber\\
g_{0i}&=&0\nonumber\\
g_{ij} &=&-\delta_{ij}\left(1+2{Gm\over r}+{G^2m^2\over r^2}\right) -{r_ir_j\over
r^2}{G^2m^2\over r^2}+\ldots\label{eq:sch}
\end{eqnarray}
as well as associated quantum corrections\cite{db2}.  The
classical correction arises from the square-root nonanalytic term
in momentum space.

Again we see then that the one-loop term contains classical (and
quantum) physics.  Despite the dimensionful coupling constant, the
key feature has again been the presence of square root nonanalytic
terms.

Classical results can also be found in other systems, not just in
energy momentum tensor form factors.  An example from
electromagnetism involves the interaction between an electric
charge and a neutral system described by an electric/magnetic
polarizability. The classical physics here is clear---the presence
of an electric charge produces an electric dipole moment $\vec{p}$
in the charge distribution of the neutral system, the size of
which is given in terms of the electric polarizability $\alpha_E$
via
\begin{equation}
\vec{p}=4\pi\alpha_E\vec{E}
\end{equation}
However, a dipole also interacts with the field, via the
energy
\begin{equation}
U=-{1\over 2}\vec{p}\cdot\vec{E}=-{1\over
2}4\pi\alpha_E\vec{E}^2
\end{equation}
Since, for a point charge $\vec{E}={e\hat{r}\over 4\pi r^2}$, there exists a simple
classical energy
\begin{equation}
U=-{1\over 2}{e^2\alpha_E\over 4\pi r^4}
\end{equation}
This result can be also be seen to arise via a simple one loop
diagram, as shown in Figure 2.  Again, for simplicity, we assume
that both systems are spinless.  The two-photon vertex associated
with the electric polarizability can be modelled in terms of a
transition to a $J^P=1^-$ intermediate state ({\it cf.} Figure 2),
yielding the Compton structure
\begin{eqnarray}
{\rm Amp}_E&=&{8\pi\over
m}\alpha_E\left[\epsilon_1\cdot\epsilon_2
P\cdot k_2 P\cdot k_1+
\epsilon_1\cdot P\epsilon_2\cdot P k_1\cdot
k_2\right.\nonumber\\
&-&\left.\epsilon_1\cdot P\epsilon_2\cdot k_1 P\cdot
k_2-\epsilon_2\cdot P\epsilon_1\cdot k_2 P\cdot k_1\right]
\end{eqnarray}
where $P={1\over 2}(p_1+p_2)$ is the mean hadron four-
momentum.
One can also include the magnetic polarizability via
transition to
a $J^P=1^+$ intermediate state, yielding
\begin{eqnarray}
{\rm Amp}_B&=&{8\pi\over
m}\alpha_B\left[\epsilon_1\cdot\epsilon_2
P\cdot k_2 P\cdot k_1+
\epsilon_1\cdot P\epsilon_2\cdot P k_1\cdot
k_2\right.\nonumber\\
&-&\left.\epsilon_1\cdot P\epsilon_2\cdot k_1 P\cdot
k_2-\epsilon_2\cdot P\epsilon_1\cdot k_2 P\cdot
k_1\right.\nonumber\\
&-&\left.k_1\cdot k_2\epsilon_1\cdot \epsilon_2 P\cdot
P+\epsilon_1\cdot k_2\epsilon_2\cdot k_1P\cdot P\right]
\end{eqnarray}
Calculating Figure 2 via standard methods, and keeping the
nonanalytic pieces of the various Feynman triangle integrals, one
finds the threshold amplitude
\begin{equation}
{\rm Amp}={e^2q^2m\over 4\pi}\left[2\alpha_E\pi^2\sqrt{M^2\over
-q^2}+({11\over 3}\alpha_E+{5\over 3}\beta_M)\right]
\end{equation}
where we have indicated the separate contributions from pole
and
seagull diagrams.  Including the normalization factor
$1/4mM$ and
Fourier transforming we find the potential energy
\begin{equation}
V_{2\gamma}(r)=-{1\over 2}{\alpha_{em}\alpha_E\over
r^4}+{\alpha_{em}(11\alpha_E+5\beta_M)\hbar\over Mr^5}+...
\end{equation}
We see again that the one loop calculation has yielded the
classical term accompanied by quantum corrections.  It should be
noted here that, although we have represented the two photon
electric/magnetic polarizability coupling in terms of a simple
contact interaction, as done by Bernabeu and Tarrach\cite{bt}, the
result is in complete agreement with a full box plus triangle
diagram calculation by Sucher and Feinberg\cite{sf1}.

\begin{figure}
\begin{center}
\epsfig{file=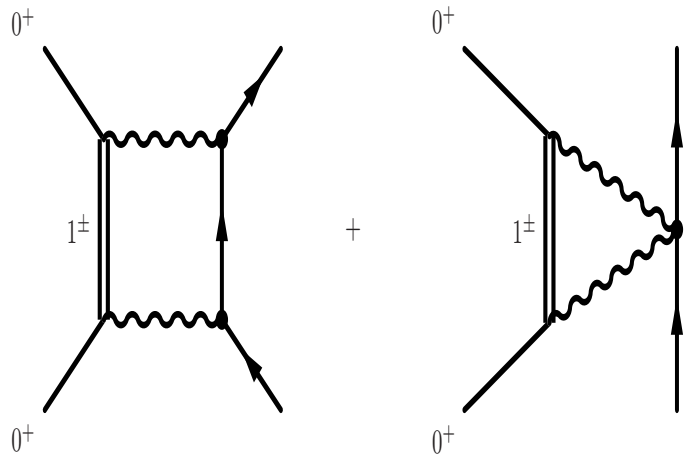,height=6cm,width=9cm} \caption{One loop diagrams used to
model the interaction of a charged particle with a neutral polarizable system. }
\end{center}
\end{figure}

There exist additional examples---a similar result obtains
by
considering the generation of an electric quadrupole moment
by an
external field gradient. Defining the field gradient via
\begin{equation}
E_{ij}={1\over 2}(\nabla_iE_j+\nabla_jE_i))
\end{equation}
and the quadrupole polarizability via
\begin{equation}
Q_{ij}=4\pi\alpha_{E2}E_{ij}
\end{equation}
The classical energy due to interaction of this moment with
the
field gradient is given by
\begin{equation}
U=-{1\over 2}\alpha_{E2}E_{ij}E_{ij}
\end{equation}
The quadrupole polarizability can be modelled in terms of excitation to a $J^P=2^+$
excited state and again, a simple one loop calculation finds a combination of
classical and quantum terms.  Similarly, in a gravitational analog, the presence of
a point mass produces a field gradient which generates a gravitational quadrupole,
which in turn interacts with the field gradient and leads to a classical energy.

Finally, the gravitational potential between two heavy masses has
been treated to one loop in an effective field theory treatment of
quantum gravity\cite{geft}. Again, the diagrams involving two
graviton propagators in a loop yield square root nonanalytic terms
which reproduce the nonlinear classical corrections to the
potential which are predicted by general
relativity\cite{donoghue}. This feature has been known for some
time\cite{gupta}.

\section{A dispersive treatment}

The lesson here is clear---these examples all involve one loop
diagrams which contain a combination of classical {\it and}
quantum mechanical effects, wherein the classical piece is
signaled by the presence of a square root nonanalyticity while the
quantum component is associated with a $\ln -q^2$ term.  These
results violate the usual expectation of the loop-$\hbar$
expansion. We can further understand the association of classical
effects with massless particles by studying a dispersive
treatment.  In this approach we can see directly that the
classical terms are associated with the dispersion integral
extending down to zero momentum, which is possible only if the
particles in the associated cut are massless.

\begin{figure}
\begin{center}
\epsfig{file=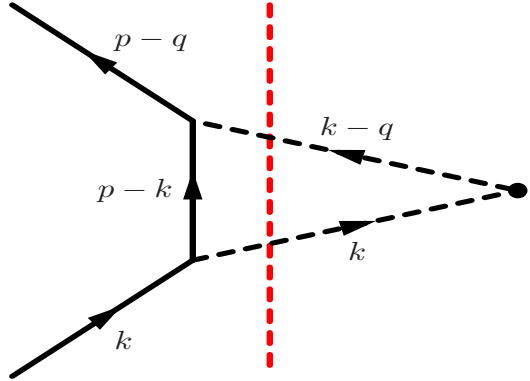,height=6cm,width=9cm} \caption{Generic triangle diagram
used in dispersive analysis.}
\end{center}
\end{figure}

It is useful to use the Cutkosky rules to look at the absorptive
component of the triangle diagram shown in Figure 3, wherein we
assume (temporarily) that the exchanged particles have mass $\mu$.
A simple calculation yields\cite{sch}
\begin{eqnarray}
&&\gamma(q^2)\equiv{\rm Abs}\int{d^4k\over (2\pi)^4}{1\over
(k^2-\mu^2)((k-q)^2-\mu^2)((k-p)^2-M^2)}\nonumber\\
&=&\int{d^4k\over (2\pi)^4}{(2\pi i)^2\delta(k^2-\mu^2)\delta((k-q)^2-\mu^2)\over
(k^2-\mu^2)((k-q)^2-\mu^2)((k-p)^2-M^2)}
\end{eqnarray}
where
\begin{equation}
\gamma(q^2)={1\over 8\pi\sqrt{q^2(4M^2-q^2)}} {\rm
tan}^{-1}{\sqrt{(q^2-4\mu^2)(4M^2-q^2)}\over q^2-2\mu^2}\label{eq:ga}
\end{equation}
 The corresponding dispersion integral is given by
\begin{equation}
\Gamma(q^2)={1\over \pi}\int_{4\mu^2}^\infty {dt\gamma(t)\over t-q^2-i\epsilon}
\end{equation}
The argument of the arctangent vanishes at threshold and the
dispersion integral yields a form of no particular interest.
On
the other hand in the limit $\mu\rightarrow 0$, the argument
of
the arctangent becomes infinite at threshold and instead we
write
\begin{equation}
\gamma(q^2)={1\over 8\pi\sqrt{q^2(4M^2-q^2)}}\left[{\pi\over 2} -{\rm
tan}^{-1}\sqrt{q^2\over (4M^2-q^2)}\right]\label{eq:g2}
\end{equation}
where we have separated the result into two components---the piece proportional to
$\pi/2$, which arises from the on-shell (delta function) piece of the mass M
propagator and the remaining terms which arise from the principal value integration.
The dispersion integral now begins at zero and yields a logarithmic result from
pieces of $\gamma(q^2)$ which behave as a constant as $q^2\rightarrow 0$, while
square root pieces arise from terms in $\gamma(q^2)$ which behave as $1/\sqrt{q^2}$
in the infrared limit.  From Eq. \ref{eq:g2} we see that the former---the quantum
component---arises from the principal value integration while the latter---the
classical component---is associated with the on- shell contributions to
$\gamma(q^2)$.  This is to be expected.  A classical contribution should arise from
the case where both initial/final {\it and} intermediate state particles are on
shell and therefore physical.

In the electromagnetic case, we can understand how such a classical term arises by
writing the Maxwell equation as

$$A_\mu={1\over \Box}j_\mu$$
Since the inverse D'Alembertian corresponds to the photon
propagator, we see that components of the triangle integral
involving the massive particle being on-shell leads to physical
values of the charge density $j_\mu$ and therefore to physical
values of the vector potential. Comparing with Eqn. \ref{eq:ga} we
see that if $\mu\neq 0$ then, there exists no possibility of a
square root term and therefore no way for classical physics to
arise.  Thus the existence of classical pieces can be traced to
the existence of two (or more!) massless propagators in the
Feynman integration.

\section{Conclusions}

We have seen above that in the presence of at least two massless
propagators, classical physics can arise from loop contributions,
in apparent contradiction to the usual loop- $\hbar$ expansion
arguments.  The presence of classical corrections are associated
with a specific nonanalytic term in momentum space. Using a
dispersion integral the origin of this phenomenon has been traced
to the infrared behavior of the Feynman diagrams involved, which
is altered dramatically when the threshold of the dispersion
integration is allowed to vanish, as can occur when two or more
massless propagators are present.  We conclude that the standard
expectation that the loop expansion is equivalent to an $\hbar$
expansion is not valid in the presence of coupling to two or more
massless particles.

\begin{center}
{\bf Acknowledgements}
\end{center}
We thank A. Zee for comments on the manuscript. This work was supported in part by
the National Science Foundation under award PHY-02-44802.

\end{document}